\documentclass{imsart}

\usepackage{amsthm,amsmath,bm,dsfont,amssymb,graphicx,multirow}
\usepackage[export]{adjustbox}
\usepackage[figuresright]{rotating}
\usepackage[colorlinks,citecolor=blue,urlcolor=blue]{hyperref}
\usepackage{natbib}

\newcommand{\indicator}[1]{\mathds{1}_{\left[ {#1} \right] }}

\begin{document}

\begin{frontmatter}

\title{Identifying overlapping terrorist cells from the Noordin Top actor\textendash event network}
\runtitle{Overlapping mixture model for terrorists network}

\author{\fnms{Saverio} \snm{Ranciati}\ead[label=e1]{saverio.ranciati2@unibo.it}\corref{cor1}\thanksref{t1}},\author{\fnms{Veronica} \snm{Vinciotti}\ead[label=e2]{veronica.vinciotti@brunel.ac.uk}}, \and\author{\fnms{Ernst C.} \snm{Wit}\ead[label=e3]{wite@usi.ch}}

\thankstext{t1}{corresponding author.}

\affiliation{$^\star$University of Bologna, Brunel University London$^\dagger$ \and University of Groningen$^\ddagger$}
\address{Saverio Ranciati \\ Department of Statistical Sciences \\ University of Bologna \\ Via delle Belle Arti 41, Bologna, 40127 \\ Italy \\ \printead{e1}}
\address{Veronica Vinciotti \\ Department of Mathematics \\ Brunel University London \\ Uxbridge UB8 3PH  \\ The UK \\ \printead{e2}}
\address{Ernst C. Wit \\ Institute of Computational Science \\ Universit\`a della Svizzera italiana \\ Via Buffi 13, CH-6904 Lugano \\ Switzerland\\ \printead{e3}}

\runauthor{Ranciati, S., Vinciotti, V., Wit, E.C.}

\begin{abstract}
Actor\textendash event data are common in sociological settings, whereby one registers the pattern of attendance of a group of social actors to a number of events. We focus on 79 members of the  Noordin Top terrorist network, who were monitored attending 45 events. The attendance or non-attendance of the terrorist to events defines the social fabric, such as group coherence and social communities. The aim of the analysis of such data is to learn about the affiliation structure. Actor\textendash event data is often transformed to actor\textendash actor data in order to be further analysed by network models, such as stochastic block models. This transformation and such analyses lead to a natural loss of information, particularly when one is interested in identifying, possibly overlapping, subgroups or communities of actors on the basis of their attendances to events.
In this paper we propose an actor\textendash event model for overlapping communities of terrorists, which simplifies interpretation of the network. We propose a mixture model with overlapping clusters for the analysis of the binary actor\textendash event network data, called {\tt manet}, and develop a Bayesian procedure for inference. After a simulation study, we show how this analysis of the terrorist network has clear interpretative advantages over the more traditional approaches of affiliation network analysis.
\end{abstract}

\begin{keyword}
\kwd{Bayesian modeling}
\kwd{mixture models}
\kwd{MCMC algorithm}
\kwd{network}
\kwd{overlapping clusters}
\end{keyword}

\end{frontmatter}

\section{Introduction}
\label{sec:Introduction}
Networks are an intuitive and a powerful way to describe interactions among individuals in many fields of application. In social sciences, for example, network structures describe concisely the observed relationships among people, tribes, social media accounts and so forth. A recent review about statistical methods and models used in this research area can be found in \cite{kolaczyk2009statistical}. Most of the literature on modelling network data can be grouped into three main branches, with some natural overlapping between the categories: stochastic block models, exponential random graph models, latent space models.  Stochastic Block Models (SBMs) date back to the work of \cite{holland1983stochastic}, where the idea of modeling partitions of the network, called blocks or communities, was first introduced. Since then,  numerous extensions, such as mixed memberships and dynamic networks, have been proposed \citep{wang1987stochastic,nowicki2001estimation,airoldi2008mixed,xing2010state}. Another way to  summarize a network structure is to model the amount of sub-structures, in a graphical and topological sense, comprising the network itself. This approach has been formulated as the exponential random graph model in the early work of \cite{frank1986markov}; see also \cite{wasserman1996logit} and  \cite{robins2007recent} for a review of some recent developments. Finally, the last framework deals with individuals in the network and their relations by projecting them into a latent space, where the probability of interaction between units is modeled based on their distance in this non-observable representation \citep{hoff2002latent}. Recent extensions of this model allow incorporating more complex features of the data, such as clustering and dynamic evolution \citep{handcock2007model,raftery2012fast,doi:10.1093/biomet/asu040,sewell2017latent}. A thorough survey on some of the most frequently used statistical network models is provided in \citet{goldenberg2010survey}.

The approaches mentioned above are mostly developed on network data where all nodes, or actors, are of the same nature. Some network data, however, are provided in the form of attendances of individuals, \emph{actors}, to \emph{events}. These data are also called two-mode networks, bipartite graphs or affiliation networks \citep[Chapter 8]{wasserman1994social}. Examples of these networks include: people visiting movies, nations belonging to alliances and co-sponsorships of legislative bills; see \cite{doreian2004generalized} for references.  There are only few models that deal directly with this actor\textendash event organization of affiliation networks. In \citet{skvoretz1999logit}, the authors cast the problem of analyzing two-mode networks in the framework of logistic regression, whereas in \citet{wang2009exponential}, affiliation network analysis with exponential random graph models is discussed. In most cases, transformation procedures are used to change \emph{actor\textendash event} data to \emph{actor\textendash actor} data. A recent example is \cite{signorelli2018penalized}, who provide a penalized approach for network data representing co-sponsorships of legislative bills in the Italian Parliament.

Transforming the data has the inherent drawbacks of information loss \citep{neal2014backbone}. In addition, in many situations, it is of prime interest to identify clusters, or \emph{communities}, of individuals within the network according to their preferences to attend specific events instead of being based on how they interact with each other. Parallel to SBMs for actor\textendash actor data, there is then the need of a clustering model for actor\textendash event data, whereby an actor (unit) is allocated into a community (cluster) based on their probability of attendance to the various events. One recent contribution is provided by \citet{aitkin2016statistical}, who propose a Rasch model approach for clustering actor\textendash event data.   Differently to their work, we expect the communities to potentially overlap with each other and we thus propose a model that allows for this. Our model is defined and parameterised in such a way that the overlap between clusters has a specific meaning, leading to parsimony and to a clear interpretation of the results. In this sense, we also depart from the literature on mixed-membership SBMs for actor\textendash actor data \citep{airoldi2008mixed}, where the SBM is extended by allowing a degree of membership for each unit to all the communities in the network.

To summarize the contribution of our work, this paper proposes a mixture model formulation that can be applied directly to actor\textendash event data in order to find communities of actors on the basis of their patterns of attendance to events. Our model accommodates for the possibility of potentially overlapping groups, and has a parsimonious formulation in terms of the number of parameters needed to represent cluster-specific probabilities of attendances to events. In particular, the parameters of the overlapping clusters are linked to the parameters of the originating clusters via a chosen function, leading to a clearer interpretation of belonging - in a `hard' clustering sense - to more than one group simultaneously.

\section{Motivating example: Noordin Top terrorist network}
\label{sec:motivating}
In this paper we consider the Noordin Top terrorist network dataset, which contains information about 79 terrorists and their activities in Indonesia and nearby areas, covering the period from 2001 to 2010 \citep{everton2012disrupting,aitkin2016statistical}. The network revolved around Noordin Mohammad Top, also known as `Moneyman', his main collaborator Azahari Husin, and their affiliates. Data were periodically collected by the \cite{ICG} in an exhaustive qualitative format. Information was later summarized  by \cite{everton2012disrupting} into relationships between terrorists, attendances to events and individual data on each terrorist, such as level of education, nationality, etc. The two-mode \emph{actor\textendash event} network focuses on the recorded attendances of the 79 terrorists to the 45 events. These events are meetings of various type.
In particular, they have been classified into: eight organizational meeting (\emph{ORG}), five operations, i.e. bombings (\emph{OPER}), eleven training events (\emph{TRAIN}), two financial meetings (\emph{FIN}), seven logistics meetings (\emph{LOGST}) and twelve events generically categorized as `meetings' (\emph{MEET}).

\begin{figure}[ht]
\centering
\includegraphics[scale=0.51]{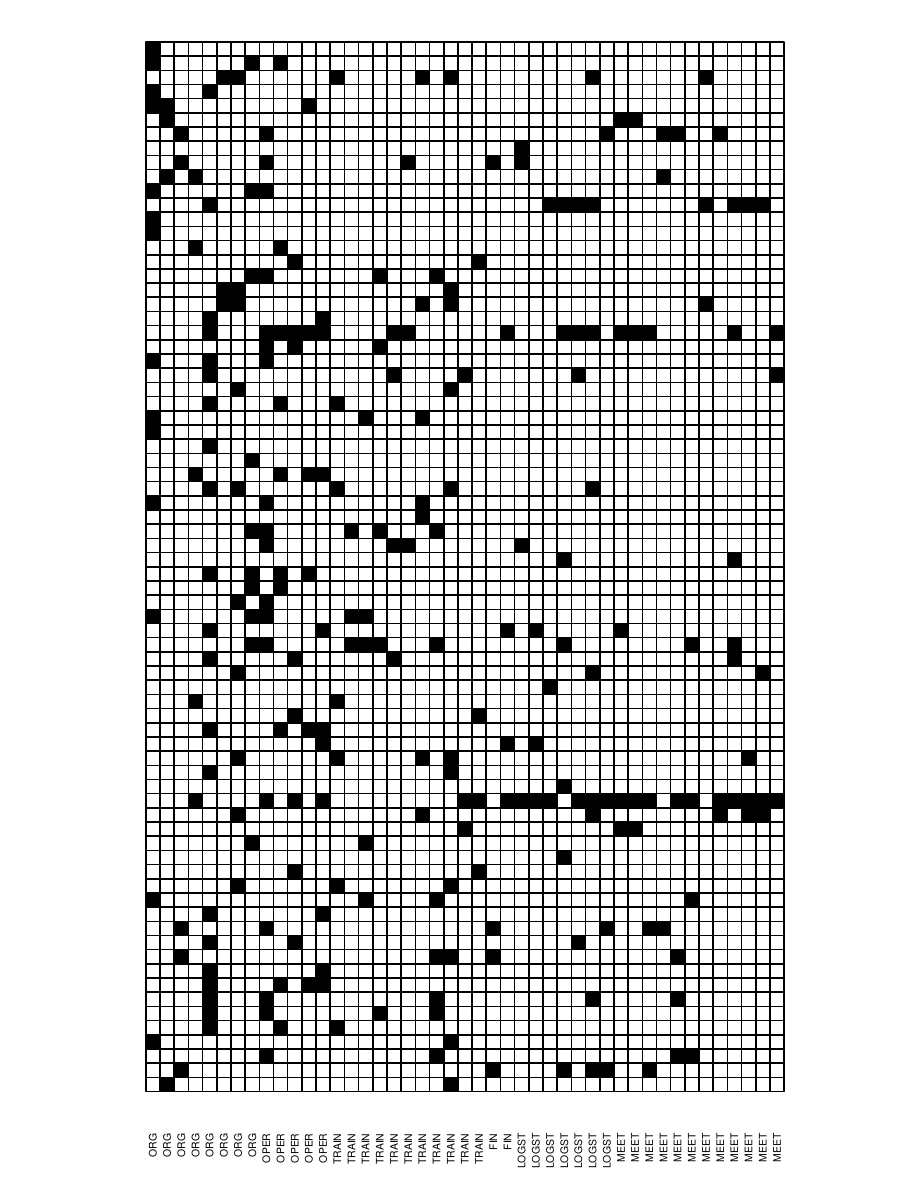}\includegraphics[scale=0.324,left]{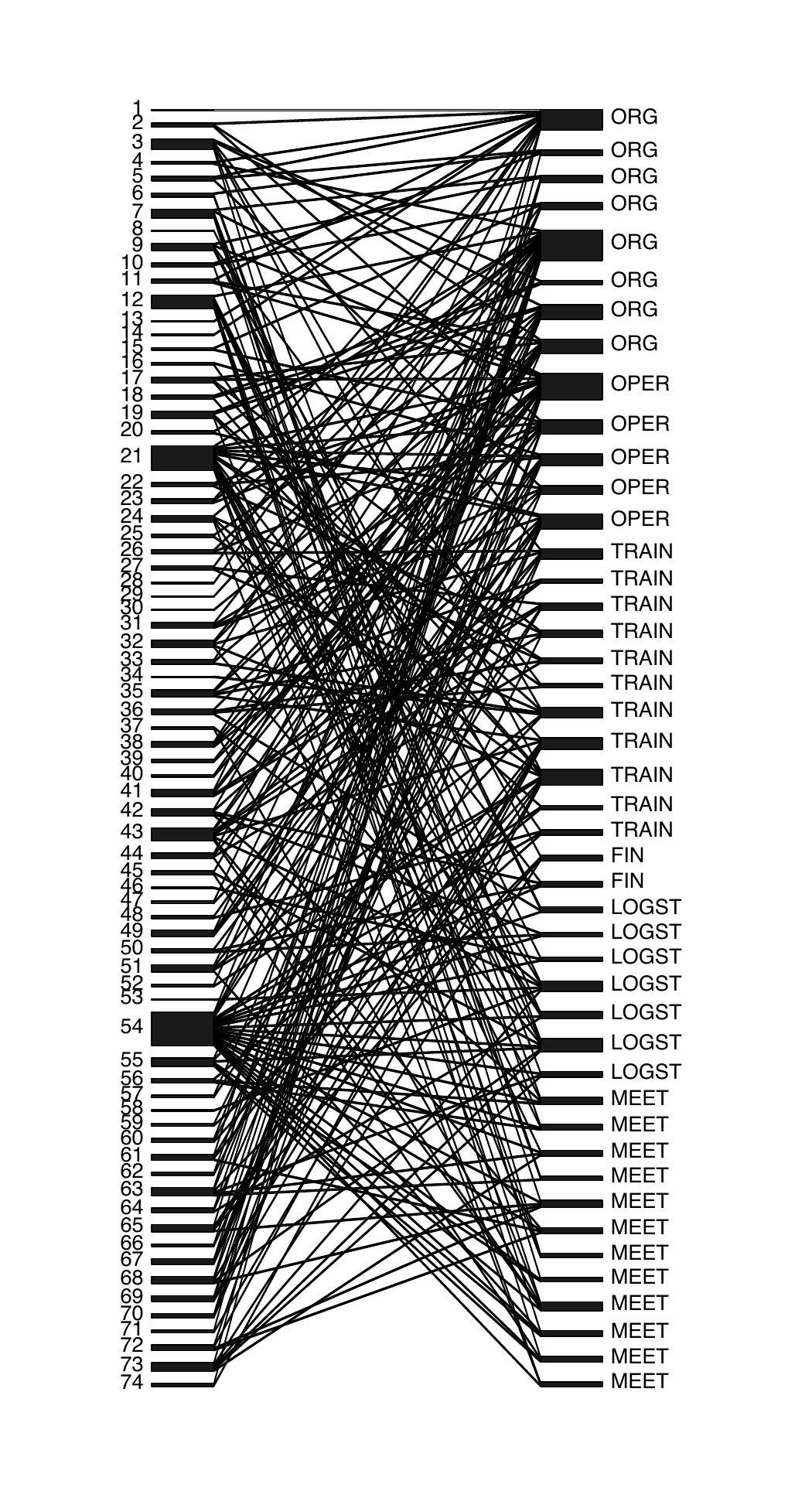}\\
~~~~~~~~~~~~(a)~~~~~~~~~~~~~~~~~~~~~~~~~~~~~~~~~~~~~~~~~~~~~~~~~~~~~~~ (b)
\centering\caption{\label{terror_fig1} \emph{(a)} Visualization of the attendances as black boxes for the 74 terrorists (rows) and the 45 events (columns). A black box depicts a connection between a terrorist and an event, while a white box indicates a terrorist not attending that event.  \emph{(b)} Visualization of the attendances as black lines. The width of the left rectangles is proportional to the connections (attendances) of each terrorist to the 45 events, whereas the width of the right rectangles is proportional to the number of terrorist attending each event. Terrorists attending no event are not visualized.}
\end{figure}

One salient feature of the network is its sparse structure, with not so many attendances recorded with respect to the total number of terrorists and events, as can be seen in Figure \ref{terror_fig1}a. Figure \ref{terror_fig1}b shows how there are some terrorists and events capitalizing most of the connections.

It is believed that a network of terrorists often operates by communities within the networks itself, whereby the individual terrorists are organized according to their role and contribution to the different activities of the whole group. More importantly, it is likely that individuals do not belong to a single community, but to more than one sub-structure in the network. The aim of this paper is to develop a model which can identify such structures (communities) among terrorists (actors) based on their patterns of attendances to the meetings (events). The proposed model can be applied to any actor\textendash event network, such as people visiting movies, nations belonging to alliances and co-sponsorships of legislative bills, when community detection on the basis of participation to the events is of interest.
\section{Model formulation}
\label{sec:model}

The driving idea is to use a model-based clustering approach to identify clusters of terrorists (actors) within the network, based on their attendances to events of different nature (bombings, trainings, financial meetings and so forth), by allowing for these communities to be potentially overlapped. We name the proposed model \emph{\textbf{m}ultiple \textbf{a}llocation model for \textbf{net}work data} (\texttt{manet}).

\subsection{Traditional model-based clustering with finite mixture model}
\label{subsec:mixture}
Data are organized in an $n \times d$ matrix of observations $y_{ij}$, pertaining to $n$ individuals and their attendances to $d$ events. Each element $y_{ij}$ is a binary random variable, with $y_{ij}=1$ if subject $i$ attends event $j$. We assume there exist $K$ sub-populations of individuals with cluster proportions $\boldsymbol{\alpha}=(\alpha_1,\dots, \alpha_K)$. In the traditional setting, where clusters are mutually exclusive, this vector satisfies the conditions (i) $\alpha_k \geq 0$, for each $k$, and (ii) $\sum_{k=1}^K \alpha_k=1$ \citep{aitkin2016statistical}.
The task is to group together units sharing the same preferential attendance to the $d$ events. Given the binary nature of response variables $y_{ij}$ and assuming independence, the marginal density of an observed attendance profile can be represented by $\mathbf{y}_i | (\boldsymbol{\alpha}, \boldsymbol{\pi}, K) \, \sim \, \sum_{k=1}^{K} \alpha_k \prod_{j=1}^d \text{Ber}\big(y_{ij}; \pi_{kj}\big)$, with $\mathbf{y}_i=(y_{i1},y_{i2},\dots,y_{ij},\dots,y_{id})$ the attendance profile of the $i$-th individual to the $d$ events and cluster specific parameters for the probability of attendance, $\pi_{kj}$, collected in $\boldsymbol{\pi}$.  A hierarchical representation is available after introducing a unit-specific latent variable $\mathbf{z}_i=(z_{i1},\dots,z_{iK})$: if unit $i$ belongs to cluster $k$, the vector is full of zeros except for the $k$-th element $z_{ik}=1$, so $\text{P}(z_{ik}=1)=\alpha_k$ and $\sum_{k=1}^K z_{ik}=1$, leading to the equivalent hierarchical conditional representation
\begin{equation*}
\mathbf{z}_i | \boldsymbol{\alpha} \, \sim \, \text{Multinom}\big(\alpha_1,\dots,\alpha_K\big), \:\:\: \mathbf{y}_i | (\mathbf{z}_i,\boldsymbol{\pi}_k) \, \sim \,  \prod_{j=1}^d \text{P}\big(y_{ij} | z_{ik}=1, \boldsymbol{\pi}_{k:z_{ik}=1} \big).
\end{equation*}
For each individual $i$, the model assumes the attendances to events $j$ and $j'$ to be independent from one another, for all $j,j'=1,\dots,d$ and $j\neq j'$.

\subsection{Multiple allocation model for network data (\texttt{manet})}
In many cases, one is interested in groups that are not mutually exclusive, allowing an actor to be allocated simultaneously to potentially more than a single cluster of the mixture model. This problem has been addressed in the statistical literature by mixture models with overlapping clusters \citep{ranciati2017mixture}. In order to cluster \emph{actor\textendash event} data by allowing possible overlaps,  we relax conditions (ii) on the proportions $\boldsymbol{\alpha}$ and the condition regarding the allocation vector, $\sum_{k=1}^K z_{ik}=1$ for each $i$. Each individual will be allowed to belong to any number of the $K$ classes. Thus, the number of all possible group membership configurations is equal to $K^{\star}=2^K$.

Instead of working with the latent variables $\boldsymbol{z}_i$, we define a new $K^{\star}$-dimensional allocation vector $\boldsymbol{z}^{\star}_i$ that satisfies $\sum_{h=1}^{K^{\star}} z^{\star}_{ih}=1$. We can establish a 1-to-1 correspondence between $\boldsymbol{z}_i$ and $\boldsymbol{z}^{\star}_i$, by introducing a $K^{\star} \times K$ binary matrix $U$, with  $z^{\star}_{ih}=\indicator{\boldsymbol{u}_h=\boldsymbol{z}_i}$, with $\boldsymbol{u}_h$ denoting the $h$-th row of $U$. For example, when $K=2$, individual $i$ may be assigned to the first cluster, $\boldsymbol{z}_i=(1,0)$, the second cluster $\boldsymbol{z}_i=(0,1)$, both of them $\boldsymbol{z}_i=(1,1)$ or none $\boldsymbol{z}_i=(0,0)$ and we have
$$U=\left( \begin{matrix}
0 & 0 \\
1 & 0 \\
0 & 1 \\
1 & 1
\end{matrix} \right).$$

We can now switch from a mixture model with $K$ overlapping \emph{parent} clusters to a finite mixture of $K^{\star}$ non-overlapping \emph{heir} clusters. Given our new assumptions on the proportions of the \emph{parent} mixture model, the model formulation changes to $$\boldsymbol{y}_i | (\boldsymbol{\alpha}^{\star}, \boldsymbol{\pi}^{\star}, K) \, \sim \, \sum_{h=1}^{K^{\star}} \alpha^{\star}_h \prod_{j=1}^d \text{Ber}\big(y_{ij}; \pi^{\star}_{hj}\big),$$ where now $\text{P}(z^{\star}_h=1)=\alpha^{\star}_h$ and $\boldsymbol{\pi}^{\star}_h$ are the attendance probabilities for the $d$ events for units whose distribution function is given by the non-overlapping cluster $h$. We specify a conjugate Dirichlet distribution for the proportions $\boldsymbol{\alpha}^{\star}$, that is $\text{P}(\boldsymbol{\alpha}^{\star} | \boldsymbol{a})=\text{Dir}(a_1, \dots, a_{K^{\star}})$. From $\boldsymbol{\alpha}^{\star}$ we can always compute back the overlapping proportions $\boldsymbol{\alpha}$ with $\alpha_k=\sum_{h=1}^{K^{\star}}\alpha^{\star}_h u_{hk}$.

In order for the overlapping mixture model to have any use and purpose, the original \emph{parent} cluster parameters should affect the \emph{heir} cluster parameters. In particular, the probability $\pi^{\star}_{hj}$ for \emph{heir} cluster $h$ of attending event $j$ should depend on the parameters $\{\pi_{kj}~|~u_{hk}=1\}$ of the \emph{parent} clusters involved in the formation of \emph{heir} cluster $h$. This can be done in a number of ways, which is described more in detail in the next paragraph.

\subsubsection*{Linking parent and heir cluster parameters}
We define the probability to attend event $j$ when belonging to heir cluster $h$ through a function $\psi \big(\boldsymbol{\pi}_j,\boldsymbol{u}_h\big) : \mathbb{R}^K \times \{0,1\}^K \rightarrow \mathbb{R}$, so that we can compute $\pi^{\star}_{hj}$ by looking at which parent clusters originated $h$, through the vector $\boldsymbol{u}_h$, and combining their corresponding probabilities $(\pi_{1j},\dots,\pi_{Kj})$.   {\color{black} By changing the definition of $\psi$ one can alter the interpretation of the  multiple allocation clusters. We argue that in many real world scenarios the minimum operator, defined by
$$\pi^{\star}_{hj}= \psi \big(\boldsymbol{\pi}_j,\boldsymbol{u}_h\big) = \left\{
      \begin{array}{rl}
			\text{min}\left\{\pi_{kj}~|~u_{hk}=1 \right\}
			& \mbox{if }\sum_{k} u_{hk} >0 \\
			0
			& \mbox{if }\sum_{k} u_{hk} =0 \\
			\end{array}
			\right.$$
is particularly sensible. Real world two-mode data, such as the Noordin Top network discussed in this paper (Section \ref{sec:terrorist}) and the Southern Women Mississippi two-mode network (Supplementary Materials), are often characterized by a sparse attendance structure and multiple allocation clusters are most naturally defined as groups of individuals that attend only those events attended by all the associated primary clusters.
For the simple case that $K=2$, an individual $i$ belonging to both clusters, $\mathbf{z}_i=(1,1)$, deciding whether to attend an event $j$ or not, will do so by following the lowest `preference' for that specific event, that is $\psi(\pi_{1j},\pi_{2j})=\text{min}(\pi_{1j},\pi_{2j})$. The multiple allocation cluster will tend to attract units that have generally a low probability of attendance to many events but a high attendance probability to a small number of events that are jointly attended by units in both primary clusters. From a Venn diagram perspective, this can be viewed as an `intersection' of parent clusters.
%From a Venn diagram perspective, we are implying multiple allocation heir clusters to be intersections of the parent clusters originating them, intersections that are however `smaller' than the parent clusters themselves. In addition, under this combining function $\psi=\min\{\cdot\}$, individuals attending few events will tend to be allocated into multiple allocation heir clusters.
In the less common scenario of dense two-mode data, it is more sensible to choose the maximum $\psi=\max\{\cdot\}$ as the operator. This will tend to allocate units with a high number of attendances into multiple allocation clusters, loosely corresponding to a union of parent clusters. %In this case, the set intersections defined by $\psi$ will usually be `bigger' than the parent clusters originating them.
}

{\color{black} As well as giving a clear meaning to the overlapping clusters and thus providing a more natural interpretation of the results, the main purpose of the link function is to reduce the number of parameters in the model.
Indeed, while we pay the price of increasing the number of proportions from $K$ to $K^{\star}$, the new quantities $\boldsymbol{\pi}^{\star}$ are not additional parameters and they can be computed from the \emph{parent} parameters $\boldsymbol{\pi}$ without increasing the parameter space's dimensionality.  This is key to the proposed model and distinguishes it from those presented in the literature, with the closest competitor being the mixed-membership SBM \citep{airoldi2008mixed} for actor\textendash actor data. Indeed, mixed-membership SBM  allows allocation to multiple clusters but there are some main differences. Firstly, the current implementation is not suited to analyzing affiliation networks (bipartite graphs). Second, mixed-membership SBM provides a form of `soft clustering', where the degree of membership reflects how strongly a unit resembles the others in the cluster: the degrees for each unit have to sum up to 1, which means that a unit cannot `strongly' -- i.e., with a high probability -- belong to more than one cluster. In our approach instead, we work with an underlying `hard clustering', thus incorporating situations not contemplated by mixed-membership SBM. In terms of number of parameters, mixed-membership SBM  requires a number of parameters proportional to $K^{\star}=2^K$, on par with a conventional (non-overlapping) mixture of Bernoulli distributions. Our model instead  allows for the overlap to be reflected in the parameter estimation, with the introduction of the $\psi(\cdot)$ function that links the parameters of the $K^{\star}$ heir clusters to those of the $K$ parent clusters, resulting in a number of parameters proportional to $K$. This not only leads to more parsimonious models but also leads to a clearer interpretation of the resulting clusters.}

\subsection{Bayesian inference}
\label{subsec:bayes}
In this section, we discuss the estimation of the parameters in our model, namely the prior membership probabilities $\boldsymbol{\alpha}$ and the probabilities of attendance to events $\boldsymbol{\pi}$.
The updated hierarchical formulation of non-overlapping mixture of the \textit{parent} clusters is given by
\begin{eqnarray*}
\text{P}(\boldsymbol{\alpha}^{\star} | \boldsymbol{a})=\text{Dir}(a_1, \dots, a_{K^{\star}}),&& \:\: \text{P}(\boldsymbol{\pi}| \boldsymbol{b}_1, \boldsymbol{b}_2)=\prod_{k=1}^{K}\prod_{j=1}^{d} \text{Beta}(\pi_{kj} ; b_{1kj}, b_{2kj})\\
\text{P}(\boldsymbol{z}^{\star}_i | \boldsymbol{\alpha}^{\star})  = \prod_{h=1}^{K^{\star}} \big( \alpha^{\star}_{h} \big)^{z^{\star}_{ih}},&& \:\: \text{P}(\boldsymbol{y}_i | \boldsymbol{z}^{\star}_i,\boldsymbol{\pi}) =  \prod_{h=1}^{K^{\star}} \prod_{j=1}^d \biggl[ \text{Ber}\big(y_{ij}; \pi^{\star}_{hj}\big) \biggl]^{z^{\star}_{ih}}.
\end{eqnarray*}
Following this structure, the joint complete data likelihood of the non-overlapping clusters model is
\begin{eqnarray*}
 \mathcal{L}( \boldsymbol{\alpha}^{\star}, \boldsymbol{\pi}; \boldsymbol{y}, \boldsymbol{z}^{\star}) &=& \prod_{i=1}^n \biggl\{ \prod_{h=1}^{K^{\star}} \biggl[ \alpha^{\star}_h \prod_{j=1}^d \text{Ber}(y_{ij}; \pi^{\star}_{hj}) \biggl]^{z^{\star}_{ih}} \biggl\}  \\
\nonumber &=&  \prod_{h=1}^{K^{\star}} \bigg(\alpha^{\star}_h\bigg)^{n^{\star}_h}  \prod_{h=1}^{K^{\star}}  \, \prod_{i:z^{\star}_i=h} \prod_{j=1}^d \text{Ber}(y_{ij}; \pi^{\star}_{hj})\\
&=&\mathcal{L}_{\boldsymbol{z}^{\star}}( \boldsymbol{\alpha}^{\star})\mathcal{L}_{\boldsymbol{y}, \boldsymbol{z}^{\star}}( \boldsymbol{\pi}),
\end{eqnarray*}
where $n^{\star}_h=\sum_{i=1}^n z^{\star}_{ih}$  and the product $ \prod_{i:z^{\star}_i=h} $ involves only units allocated to cluster $h$. The second term, $\mathcal{L}_{\boldsymbol{y}, \boldsymbol{z}^{\star}}( \boldsymbol{\pi})$, is a function of the parameters $\boldsymbol{\pi}$ through the computed quantities $\boldsymbol{\pi}^{\star}$. In order to devise a Gibbs sampler for $\boldsymbol{\pi}$, we consider the equivalent representation for the overlapping-clusters mixture, as a function of the original \emph{parent} parameters, that is $\mathcal{L}( \boldsymbol{\alpha}^{\star}, \boldsymbol{\pi}; \boldsymbol{y}, \boldsymbol{z})$. The first term is equivalent in both parametrization thanks to the 1-to-1 correspondence between $\boldsymbol{z}$ and $\boldsymbol{z}^{\star}$, and the computability of $\boldsymbol{\alpha}$ from $\boldsymbol{\alpha}^{\star}$. We focus now on the second term of the factorization, $\mathcal{L}_{\boldsymbol{y}, \boldsymbol{z}}( \boldsymbol{\pi})$, as it is not immediately straightforward to define an equivalence. We introduce a new quantity $\boldsymbol{s}(\boldsymbol{z}_i,\boldsymbol{\pi})=\boldsymbol{s}^{(j)}_{i}$, whereby $\boldsymbol{s}^{(j)}_{i}=\boldsymbol{z}_i$ if $\sum_{k=1}^K z_{ik}=1$, whereas, if $\sum_{k=1}^K z_{ik}>1$ and if we use the minimum operator, i.e. $\psi = \min(~\cdot~)$, then $\boldsymbol{s}^{(j)}_{i}$ is a $K$-dimensional vector of zeros except for $s_{ik_{\text{min},j}}=1$, with $k_{\text{min},j}$ denoting the cluster with the lowest value among all the parameters $\boldsymbol{\pi}_k$ for a fixed event $j$. In other words, if a unit $i$ belongs to only one cluster (let us say, $k$) it will fully contribute to the posterior of the corresponding $\pi_{kj}$; but, if the unit $i$ is allocated into more than one group its contribution will be given only to the lowest parameter $\pi_{k_{\text{min},j}}$ among
all the relevant attendance probabilities $\{\pi_{kj} ~|~u_{h(i)k}=1\}$ for that $j$-th event.
This definition is compatible with the minimum operator $\psi$. For other operators, one needs to consider other solutions.

This leads to a convenient factorization of the complete data likelihood of the mixture in the $K$ space:
\begin{eqnarray*}\label{complete}
\mathcal{L}( \boldsymbol{\pi}, \boldsymbol{s}; \boldsymbol{y}, \boldsymbol{z}) =  \prod_{k=1}^{K} \prod_{j=1}^d \pi_{kj}^{\sum_{i=1}^n y_{ij}s^{(j)}_{ik}} (1-\pi_{kj})^{\sum_{i=1}^n s^{(j)}_{ik}-\sum_{i=1}^n y_{ij}s^{(j)}_{ik}}.
\end{eqnarray*}
A sketch of our sampling scheme is the following.  For each unit $i$ and \emph{heir} cluster $h$, we compute the posterior probabilities of allocation conditional on the observations and other parameters, according to
$$ \text{P}(\boldsymbol{z}^{\star}_i=h | \boldsymbol{y}, \boldsymbol{\alpha}^{\star}, \boldsymbol{\pi})= \frac{\alpha^{\star}_h \prod_{j=1}^d \text{Ber}(y_{ij}; \pi^{\star}_{hj})}{\sum_{h'=1}^{K^{\star}} \alpha^{\star}_{h'} \prod_{j=1}^d \text{Ber}(y_{ij}; \pi^{\star}_{h'j})},$$
and we sample new latent allocation values for $\boldsymbol{z}^{\star}_i$. The proportions $\boldsymbol{\alpha}^{\star}$ are updated through the corresponding full conditional distribution, $ \boldsymbol{\alpha^{\star}} \sim \text{Dir}\big(n^{\star}_1+a_{1}, \dots, n^{\star}_{K^{\star}}+a_{K^{\star}}  \big)$. Thanks to the prior-likelihood conjugacy, each of the $\pi_{kj}$ are updated via a Gibbs sampler with $$ \pi_{kj} \sim \text{Beta}\biggl(\sum_{i=1}^n y_{ij} s^{(j)}_{ik}+b_{1kj}; \sum_{i=1}^n s^{(j)}_{ik} -\sum_{i=1}^n y_{ij} s^{(j)}_{ik}+b_{2kj} \biggl).$$ We implement all the samplers in an MCMC algorithm. The latter is also part of the \texttt{R} package \texttt{manet}, available on \texttt{CRAN}.

\subsection{Selecting the number of clusters and criterion to allocate units}
\label{subsec:dic_allocation}
We select the Deviance Information Criterion \citep[DIC]{spiegelhalter2002bayesian} as the model selection criterion. This criterion has the property of being the large sample (robust) version of the AIC \citep[Ch. 3.5]{claeskens2008model}. In the DIC, two quantities are balanced, namely the goodness-of-fit and the complexity of the model. In this paper, we rely on the version DIC$_3$ proposed in \cite{celeux2006deviance}, as the original version does not deal properly with latent variables:
$$\text{DIC}(K)=-4\text{E}_{\boldsymbol{\alpha}^{\star},\boldsymbol{\pi}}[\log \text{P}(\boldsymbol{y}|\boldsymbol{\alpha}^{\star},\boldsymbol{\pi})]+2\log \hat{\text{P}}(\boldsymbol{y}),$$
where both terms can be computed starting from the values sampled at each iteration $t=1,\dots,T$ of the MCMC algorithm. In particular, $$\text{E}_{\boldsymbol{\alpha}^{\star},\boldsymbol{\pi}}[\log \text{P}(\boldsymbol{y}|\boldsymbol{\alpha}^{\star},\boldsymbol{\pi})]=\frac{1}{T} \sum_{t=1}^T \sum_{i=1}^n \log \biggl\{ \sum_{h=1}^{K^{\star}} \alpha^{\star^{(t)}}_h \prod_{j=1}^d \text{Ber}\big(y_{ij}; \pi^{\star^{(t)}}_{hj}\big) \biggl\},$$ and $$\hat{\text{P}}(\boldsymbol{y})=\prod_{i=1}^n \hat{\text{P}}(\boldsymbol{y}_i) \mbox{, where } \hat{\text{P}}(\boldsymbol{y}_i)=\frac{1}{T} \sum_{t=1}^T \biggl\{ \sum_{h=1}^{K^{\star}} \alpha^{\star^{(t)}}_h \prod_{j=1}^d \text{Ber}\big(y_{ij}; \pi^{\star^{(t)}}_{hj}\big) \biggl\}.$$ In a set of competing models, differing from one another only by $K$, we select the one with the lowest associated DIC$(K)$ value.

After the choice of $K$ and, implicitly, ${K}^{\star}$, units are allocated into clusters according to their average posterior probabilities and using the Maximum-A-Posteriori (MAP) rule. That is, individual $i$ will be assigned to cluster $h$ showing the highest value for $\bar{\text{P}}(\boldsymbol{z}^{\star}_i=h | \boldsymbol{y}, \boldsymbol{\alpha},\boldsymbol{\pi})= T^{-1} \sum_{t=1}^{T}\text{P}(\boldsymbol{z}^{\star}_i=h | \boldsymbol{y},\boldsymbol{\alpha}^{\star^{(t)}}  \boldsymbol{\pi}^{(t)})$, computed after the initial burn-in window.

\subsection{Quantifying clustering uncertainty}
\label{subsec:pcm}
As a measure of uncertainty about the clustering provided by the algorithm, we define a quantity called Posterior Confusion Matrix ({PCM}), whose entry PCM$_{hk}$ stands for the average number of actors with maximum posterior allocation for cluster $h$ that will be allocated to cluster $k$. The PCM is a non-symmetrical $K^{\star} \times K^{\star}$ matrix and is computed as follows. For each MCMC iteration $t=1,\dots,T$ and summed across all units $i=1,\dots,n$, we do the following steps:
\begin{enumerate}
\item Order the posterior probabilities $\text{P}(\boldsymbol{z}^{\star}_i=h | \boldsymbol{y}, \boldsymbol{\alpha}^{\star^{(t)}}  \boldsymbol{\pi}^{(t)})$ from highest to lowest, and collect them in a vector $\boldsymbol{\tau}_{i}^{(t)}$;
\item Define $\mathbf{r}^{(t)}_i$ as the vector of cluster labels associated to $\boldsymbol{\tau}_{i}^{(t)}$, so that $r^{(t)}_{i,1}$ is the label of the cluster with highest posterior probability (which is $\tau^{(t)}_{i,1}$) for unit $i$ at iteration $t$ among all the $K^{\star}$ possible ones;
\item Add posterior probability $\tau^{(t)}_{i,1}$ to the PCM at position $(r_{i,1}^{(t)},r_{i,1}^{(t)})$, so that the diagonal element of the matrix account for the first choice of allocation of unit $i$ at iteration $t$;
\item While keeping row $r_{i,1}^{(t)}$ fixed as a pivotal quantity of this step, add the remaining probabilities $\tau^{(t)}_{i,2},\tau^{(t)}_{i,3},\dots,\tau^{(t)}_{i,K^{\star}}$ to the corresponding positions in the PCM matrix $(r_{i,1}^{(t)}, r_{i,2}^{(t)}), (r_{i,1}^{(t)},r_{i,3}^{(t)}), \dots, (r_{i,1}^{(t)},r_{i,K^{\star}}^{(t)})$.
\end{enumerate}
To average the cumulative sums at each position of the matrix, we divide the PCM by the total number of MCMC iterations $T$. The non-rescaled version of the matrix has row sums equal to the number of units in each corresponding cluster. When rescaled by these row sums, the benchmark matrix for comparison is the identity matrix of order $K^{\star}$, corresponding to a situation with no uncertainty in the classification.

{\color{black}A well-known issue of mixture models in the Bayesian paradigm is the so-called ``label switching'' problem: that is, the likelihood of a mixture model is symmetrical with respect to permutation of the clusters' labels. This trait is inherited by the posterior distribution, unless specific constraint are applied to the prior, for example, in order to break the symmetry, but in general the resulting posterior density will have $K!$ different modes. Although a sampler should be encouraged to visit all the potential high-density regions of the posterior, in practice the MCMC chains could jump unpredictably between the modes and thus hindering the computation of summaries such as posterior means and posterior standard deviations. Many authors, in the literature, have studied this specific issue: for a review of some techniques to deal with label switching, we refer the reader to \cite{stephens2000dealing}. }

\section{Simulation study}
\label{sec:simulation}
In this section, we perform a simulation study where we compare the following algorithms: (i) the proposed model, \texttt{manet}, which uses a finite mixture of Bernoulli distributions with overlapping components (as implemented in the package \texttt{manet});  (ii) a finite mixture model of Bernoulli distribution with $K=K^{\star}$ non-overlapping components, named \texttt{mixtbern}, (iii) a variational method implementing the MixNet model of \cite{daudin2008mixture}, implemented in the R package \texttt{mixer}, which is a special case of the binary SBM proposed by \cite{nowicki2001estimation} and (iv) \texttt{blockmodels}, proposed by \cite{leger2015blockmodels}.

To measure the performance of the four models we apply the MAP rule to the estimated probabilities of allocation and we cluster units accordingly. After the classification is performed,  we compute the average misclassification error rate and the adjusted Rand index \citep{rand1971objective} for each of the four models across the independently replicated datasets. The misclassification error rate measures the fraction of units wrongly allocated with respect to the true allocations used to generated the data, whereas the adjusted Rand index (ARI) is a measure between 0 and 1 representing similarity between two different clustering, where we take one of the two to be the true allocation in the data.

\subsection{Synthetic data generated from \texttt{manet}}
\label{subsec:classification}
For the scenarios considered in this section, we generate data according to our model with varying values for the number of actors $n$ and the number of events $d$. We consider $K=3$ (i.e. $K^{\star}=8$) and set the components weights to be $\boldsymbol{\alpha}^{\star}=(0.1,0.25,0.20,0.1,0.15,0.1,0.05,0.05)$. We set the probabilities of attendances for the first event equal to $\boldsymbol{\pi}_{\cdot1}=(0.2,0.5,0.9)$ and we define the remaining vectors to be all the possible $(K!-1)$ permutations of the values in $\boldsymbol{\pi}_{\cdot1}$, by stacking the same values a number of times depending on the value of $d$ chosen.

Since \texttt{blockmodels} and \texttt{mixer} only work on actor\textendash actor data, for these two methods we transform the data to this structure by calculating the number of events attended by any two actors.  This is sufficient for \texttt{blockmodels}, which accounts for weighted edges. Since \texttt{mixer} requires a binary input, we further dichotomize the network by setting a cutoff on the number of events. For this, we select the threshold that leads to the best results for each of the methods.

\subsubsection{Classification performance}

For this simulation, we set $n=300$ and consider three possible values for the number of events, namely $d=\{6,18,38\}$. For each of the three values of $d$, we generate 25 independent datasets. We then run the algorithm by setting the true number of clusters, i.e. $K=3$ for our model or $K^{\star}=8$ for the competitors. Table \ref{tab_clasf} reports the results of this simulation in terms of the ability of allocating the actors into the 8 \emph{heir} clusters. In each sub-group defined by the value of $d$, our model achieves simultaneously lower (better) average misclassification error rate and higher  (better) average adjusted Rand index with respect to the other competitors. The closest in terms of performance is \texttt{mixtbern}, which however exhibits less stability. It is worth noticing that as the number of events, $d$, increases so does the performance improvement in the classification task: this is true for all the models with the exception of \texttt{mixer}. The loss of performance for models \texttt{blockmodels} and \texttt{mixer} is partially expected due to the loss of information after transformation of the data into a one-mode network.

\begin{table}[h!]
\begin{tabular}{|c|cccc|}
\hline
\multicolumn{5}{|c|}{\textbf{Misclassification error rate (in \%)}}\\
\hline
\multirow{2}{*}{\textit{Num. of events}} & \multicolumn{2}{|c|}{\textit{actor\textendash actor}} & \multicolumn{2}{|c|}{\textit{actor\textendash event}} \\
\cline{2-5}
& \texttt{mixtbern} & \texttt{manet} & \texttt{mixer} & \texttt{blockmodels}  \\
\hline
$d=6$ &42.67 \tiny{(5.96)} & \textbf{35.05 \tiny{(3.99)}}  & 52.16 \tiny{(2.23)}  &55.49 \tiny{(3.11)}  \\
$d=18$ & 20.89 \tiny{(2.97)} & \textbf{15.33 \tiny{(2.42)}}  & 46.89 \tiny{(5.87)} & 43.07 \tiny{(4.49)}   \\
$d=36$ & 13.67 \tiny{(4.14)}  & \textbf{6.91 \tiny{(1.53)}}   & 54.32 \tiny{(7.32)} &  30.28 \tiny{(4.76)}\\
\hline
\noalign{\vskip 2mm}
\hline
\multicolumn{5}{|c|}{\textbf{Adjusted Rand index} ($\text{ARI}_{\text{max}}=1$)}\\
\hline
\multirow{2}{*}{\textit{Num. of events}} & \multicolumn{2}{|c|}{\textit{actor\textendash actor}} & \multicolumn{2}{|c|}{\textit{actor\textendash event}} \\
\cline{2-5}
& \texttt{mixtbern} & \texttt{manet} & \texttt{mixer} & \texttt{blockmodels}  \\
\hline
$d=6$ & 0.34 \tiny{(0.08)} & \textbf{0.45 \tiny{(0.06)}}  & 0.15 \tiny{(0.03)} & 0.22 \tiny{(0.04)} \\
$d=18$ & 0.73 \tiny{(0.05)} & \textbf{0.79 \tiny{(0.04)}} & 0.31 \tiny{(0.08)} & 0.40 \tiny{(0.06)} \\
$d=36$ & 0.85 \tiny{(0.05)} & \textbf{0.93 \tiny{(0.02)}} & 0.27 \tiny{(0.08)} & 0.60 \tiny{(0.06)} \\
\hline
\end{tabular}
\caption{\label{tab_clasf}Misclassification error rate and adjusted Rand index, averaged over 25 replicated datasets, for three values of $d=\{6,18,36\}$ and four competing models; standard errors are reported between brackets. Models are categorized on the type of structure they analyze (\textit{actor\textendash actor} or \textit{actor\textendash event}); best results are highlighted in bold.}
\end{table}

\subsubsection{Convergence of parameters' posterior distributions}
\label{subsubsec:parameters}
\begin{sidewaysfigure}[p!]
\includegraphics[scale=0.5,center]{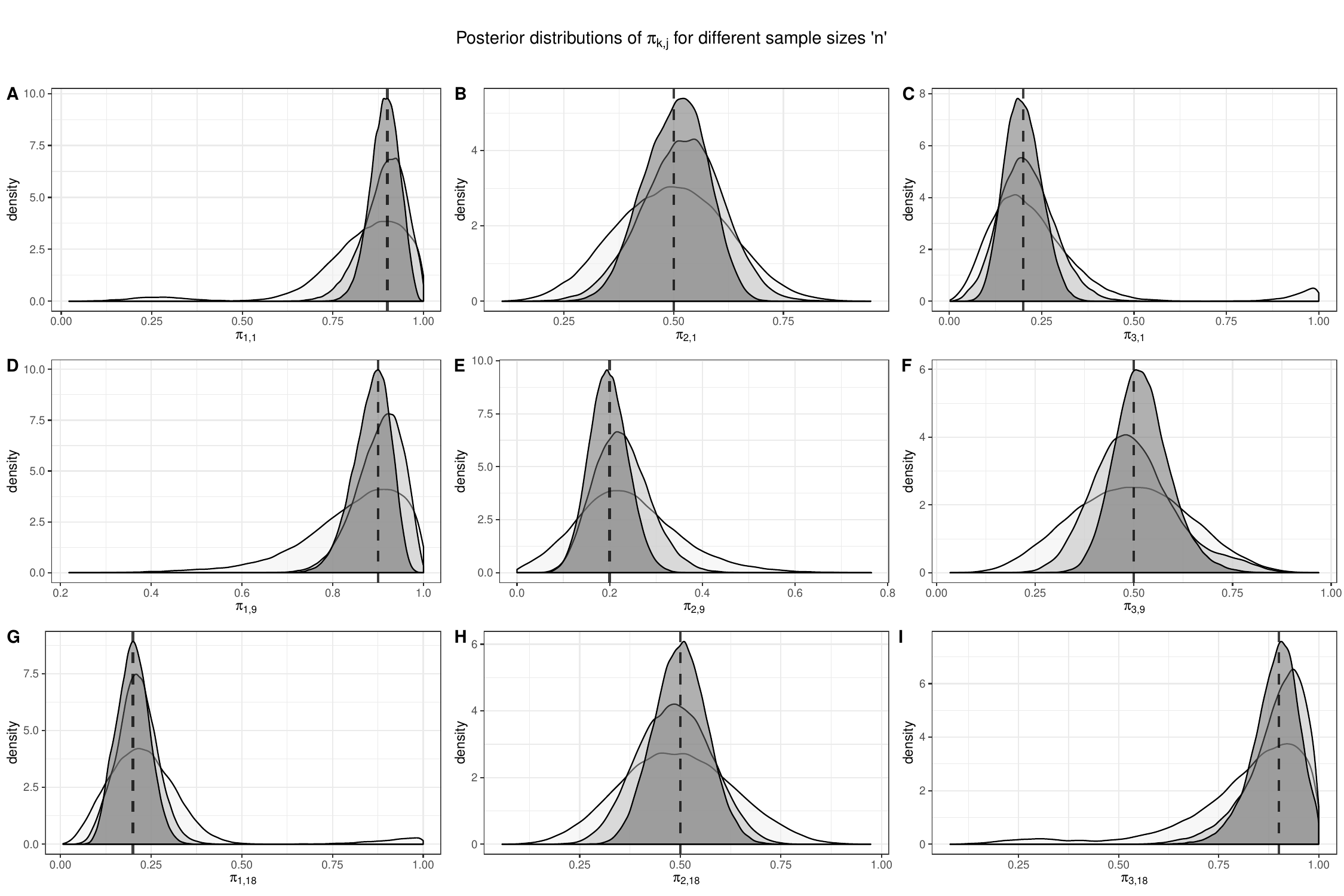}
\centering\caption{\label{plots_grid} Posterior distributions of $\pi_{k,j}$ for three events $j=\{1,9,18\}$ and all the clusters, for varying sample sizes $n=\{100,250,500\}$. Each curve collects all posterior samples (after burn-in) from the 25 replicated datasets. The grey shading in the plots (A-I) is: \textbf{white}, $n=100$; \textbf{light grey}, $n=250$; \textbf{medium grey}, $n=500$. True values of $\pi_{k,j}$ are reported as dashed vertical lines.}
\end{sidewaysfigure}

For this simulation, we focus on the convergence behavior of the posterior distributions of the attendance probabilities $\pi_{kj}$ to the true values of the data generating model. In particular, we use a fixed setting with $K=3$, $d=18$, letting the sample size vary as $n=\{100,250,500\}$. We set the true values for the $\{\pi_{kj}\}$ as described in Section \ref{subsec:classification}. For each sample size, we simulate 25 replicated datasets and we collect all posterior samples (after burn-in) of the same $n$ from each MCMC into one single chain. While this inevitably introduces some additional Monte Carlo error, the increased amount of available information should dampen this aggregation effect. Results are visualized in Figure \ref{plots_grid}.
%\begin{sidewaysfigure}[p!]
%\includegraphics[scale=0.45,center]{plots_grid}
%\centering\caption{\label{plots_grid} Posterior distributions of $\pi_{k,j}$ for three events $j=\{1,9,18\}$ and all the clusters, for varying sample sizes $n=\{100,250,500\}$. Each curve collects all posterior samples (after burn-in) from the 25 replicated datasets. The color coding in the plots (A-I) is: \textbf{red}, $n=100$; \textbf{green}, $n=250$; \textbf{blue}, $n=500$. True values of $\pi_{k,j}$ are reported as dashed vertical lines.}
%\end{sidewaysfigure}
Rows of the plot correspond to events (specifically, we are reporting $j=\{1,9,18\}$) and columns to the attendance probabilities of those events for the three different primary clusters. As expected, with increasing sample size (from $n=100$, red curve, to $n=500$, blue curve), the posterior distribution exhibits less variability, contracting around the true value, i.e., the vertical dashed line, used for the simulations. The same behavior is observed for the posterior distributions of the other $\pi_{kj}$ and the posterior distribution of $\boldsymbol{\alpha}^{\star}$, the proportions of the mixture model (not shown).

\subsubsection{Accuracy of model selection criterion}
\label{subsubsec:modelsel}

To show the behaviour of the DIC selection criterion discussed in Section \ref{subsec:dic_allocation}, we simulate 25 replicated datasets with the following configuration: $K_{\text{true}}=3$, $d=18$, increasing sample sizes $n=\{25,75,150,300\}$. For each dataset, we run the algorithm and provide three different values of $K=\{2,3,4\}$. We compute the corresponding DIC values and select the value of $K$ that achieves the lowest one. When $n=25$, we select $\hat{K}=K_{\text{true}}=3$ in 80\% of the replicated datasets; for the remaining sample sizes ($n=\{75,150,300\}$), the DIC achieves its lowest value with $\hat{K}=K_{\text{true}}=3$ in all the datasets.

\subsection{Synthetic data generated from a misspecified model}
\label{subsec:mispec}
\begin{figure}[ht!]
\includegraphics[scale=0.35,center]{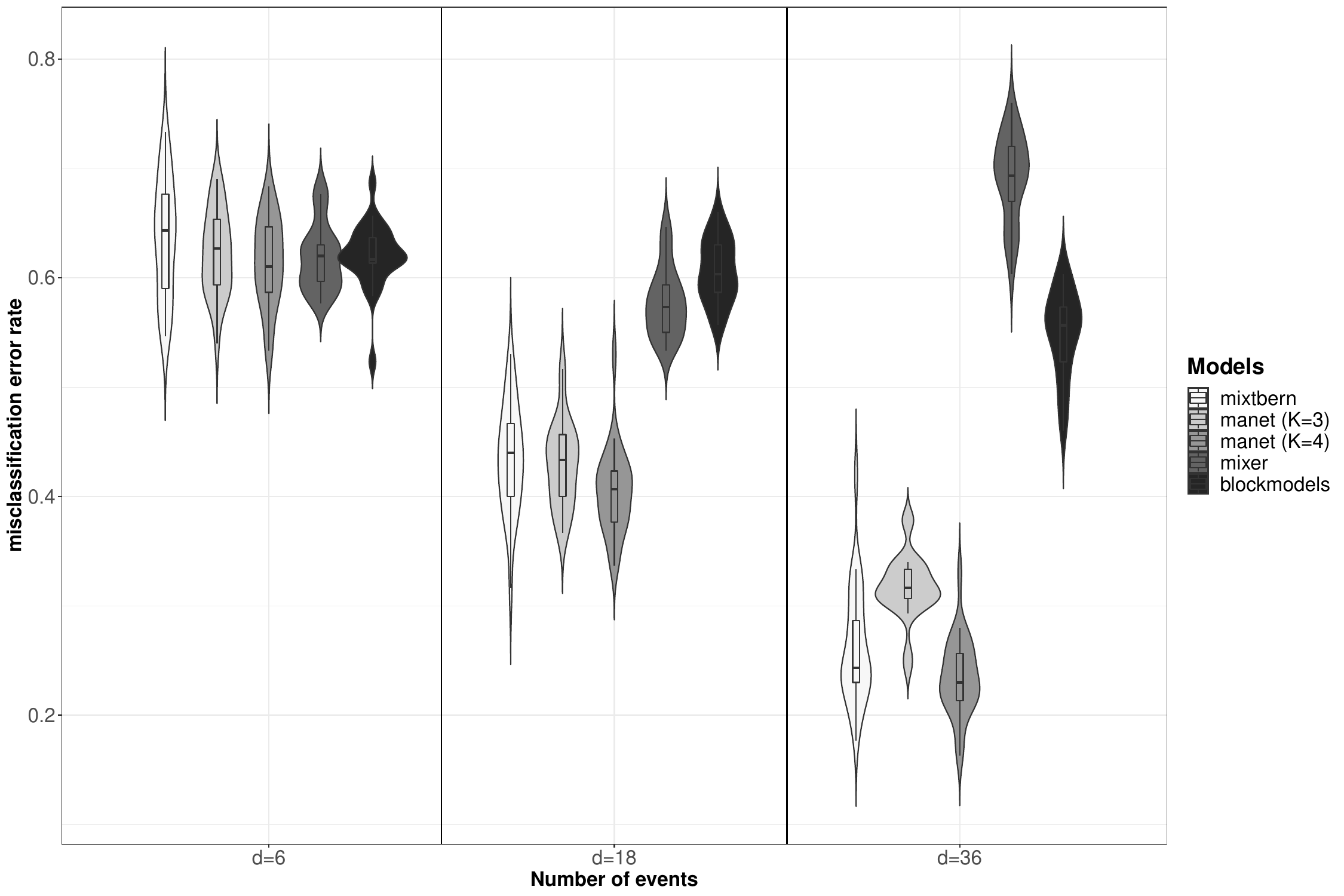}
\includegraphics[scale=0.35,center]{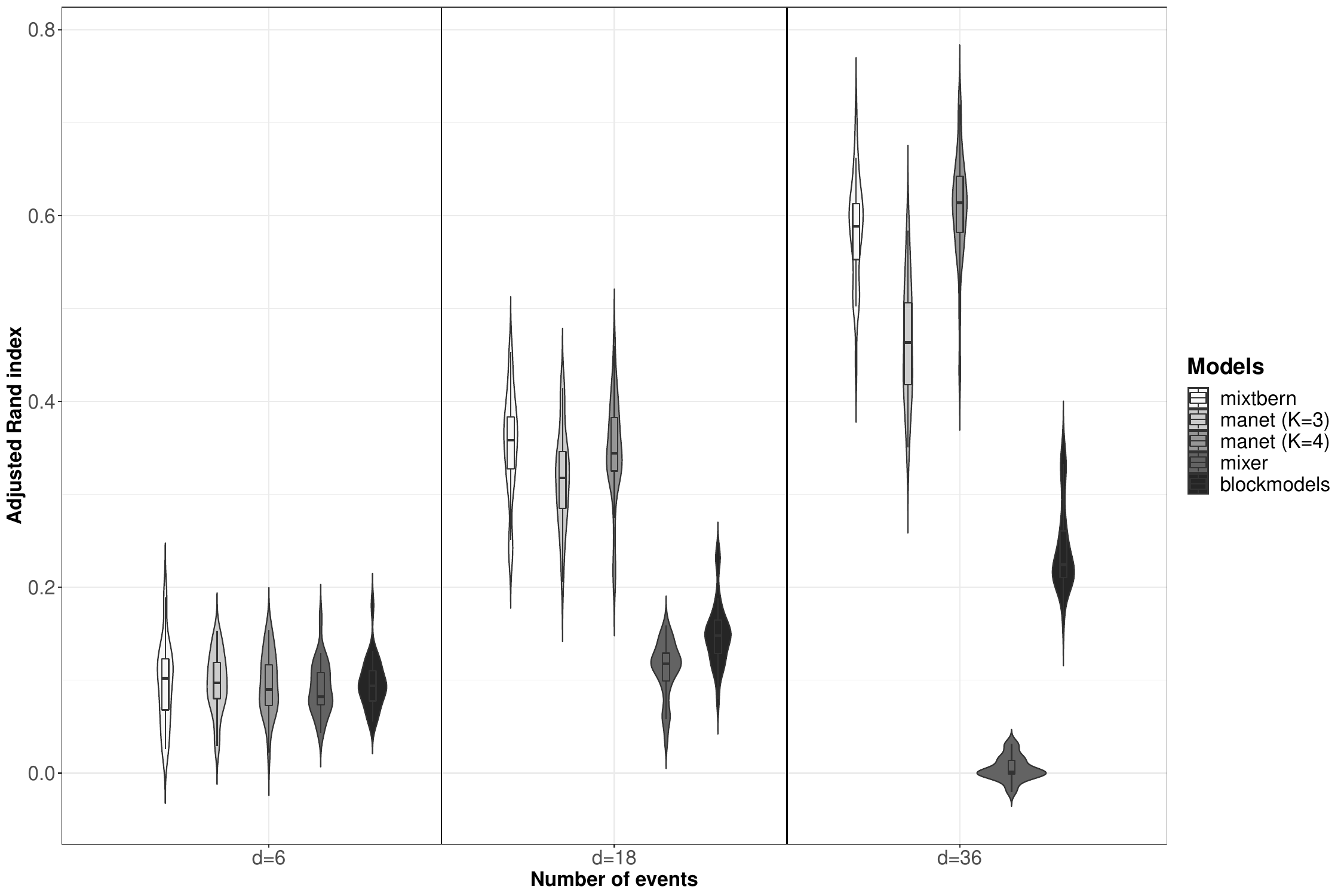}
\centering\caption{\label{misspec}Results for the simulation study on data from a misspecified model and 3 different values for the total number of events $d=\{6,18,36\}$. The violin plots report misclassification error rate (\emph{top}) and ARI (\emph{bottom}) across the 25 replicated datasets. For each value of $d$, the five boxplots in each section refer to (from left to right):  \texttt{mixtbern}, \texttt{manet} ($K=3$), \texttt{manet} ($K=4$), \texttt{mixer},  \texttt{blockmodels}.}
\end{figure}

 The previous section showed simulations on data generated by our proposed model. For the scenarios considered in this section, we consider misspecified cases. In particular, we simulate attendances for $n=300$ units to $d$ events, where $d=\{6,18,36\}$, from a mixture of independent Bernoulli distributions (\texttt{mixtbern}) with $K=8$ non-overlapping components. %In this way, we test how a model that allows for overlapping clusters performs in the degenerate case of no overlapping.
The weights for the mixture are set to $\boldsymbol{\alpha}=(0.1,0.25,0.20,0.1,0.15,0.1,0.05,0.05)$, while the probabilities of attendances $\{\pi_{kj}\}$ are defined as follows:
\begin{itemize}
\item $\boldsymbol{\pi}_{1\cdot}=(0.9,0.8,0.7,0.6,0.5,0.1)$;
\item $\boldsymbol{\pi}_{2\cdot}=(0.3,0.2,0.1,0.9,0.3,0.2)$;
\item $\boldsymbol{\pi}_{3\cdot}=(0.7,0.6,0.5,0.9,0.3,0.2)$;
\item $\boldsymbol{\pi}_{4\cdot}=(0.2,0.1,0.7,0.6,0.3,0.1)$;
\item $\boldsymbol{\pi}_{5\cdot}=(0.2,0.1,0.9,0.8,0.3,0.6)$;
\item $\boldsymbol{\pi}_{6\cdot}=(0.4,0.5,0.5,0.7,0.3,0.1)$;
\item $\boldsymbol{\pi}_{7\cdot}=(0.3,0.2,0.1,0.9,0.8,0.7)$;
\item $\boldsymbol{\pi}_{8\cdot}=(0.4,0.5,0.6,0.7,0.8,0.1)$.
\end{itemize}
The results, in terms of misclassification error rate (MCR) and Adjusted Rand Index (ARI), are visualized in Figure \ref{misspec}.
%\begin{figure}[ht!]
%\includegraphics[scale=0.4,center]{misspec1}
%\includegraphics[scale=0.4,center]{misspec2}
%\centering\caption{\label{misspec}Results for the simulation study on data from a misspecified model and 3 different values for the total number of events $d=\{6,18,36\}$. The violin plots report misclassification error rate (\emph{top}) and ARI (\emph{bottom}) across the 25 replicated datasets. For each value of $d$, the five boxplots in each section refer to (from left to right):  \texttt{mixtbern}, \texttt{manet} ($K=3$), \texttt{manet} ($K=4$), \texttt{mixer},  \texttt{blockmodels}.}
%\end{figure}

The figure is separated into three blocks, corresponding to the number of events ($d=\{6,18,36\}$). The plots are vertically separated according to the two measures of performance, MCR and ARI, respectively, which are computed on 25 replicated datasets and for five competing models: \texttt{blockmodels},  \texttt{mixer},  \texttt{mixtbern}, and  \texttt{manet} with $K=3 \rightarrow K^{\star}=8$ and $K=4 \rightarrow K^{\star}=16$ clusters. When $d=6$ all models exhibit poor performance, which is due to the difficult clustering task posed by the small number of events. For $d=18$, the true model \texttt{mixtbern} and \texttt{manet} (both $K=3$ and $K=4$) show lower error rates for the classification and a better agreement with the true cluster labels. In the scenario where $d=36$, \texttt{manet} with $K=3$ clusters performs worse than \texttt{mixtbern}. However, if we fit \texttt{manet} with $K=4$, the model has enough flexibility to accommodate 8 non-empty clusters, while being (more) parsimonious in the number of estimated parameters than \texttt{mixtbern}, and thus allowing it to perform on par with -- if not slightly better than -- \texttt{mixtbern}.

\subsection{Computational times and storage}

{\color{black}At the current stage of the implementation of our proposed method in the \texttt{R} package \texttt{manet}, the algorithm requires to store: (i) posterior probabilities of allocation $p(z | y, \dots)$ at each MCMC iteration in an $n \times K^{\star}$ matrix; (ii) components' weights in a vector of length $K^{\star}$; (iii) sampled probabilities of attendances $\pi_{kj}$ in $K$ vectors of length $d$. Among these quantities, only (i) and (ii) scale with $K^{\star}$. As far as execution times are concerned, the computational burden scales exponentially in the number of parent clusters $K$ only for the sampling of the components weights, whereas it is linear in terms of $K$ and $d$ because of the parsimonious formulation of the model. Nevertheless, we generally expect the number of overlapping cluster $K$ to be rather small in most applications, implying no need to run \texttt{manet} with large $K$ and thus longer CPU times. To provide a numerical comparison, Table \ref{comp_t} reports the computational times for the scenarios explored in the simulation studies (Section \ref{subsec:mispec}) as milliseconds per iteration (mpi), that is the execution time in seconds divided by the number of MCMC iterations.

\begin{table}[ht!]
\centering
\begin{tabular}{c|c|c|c|c}
\hline
\multirow{2}{*}{sample size $n$} & \multirow{2}{*}{\# clusters $K$} & \multirow{2}{*}{\# of events $d$} & \multicolumn{2}{c}{execution times}\\
\cline{4-5}
 & & & total elapsed time & mpi \\
\hline
\multirow{6}{*}{$300$} & \multirow{3}{*}{$3$} & 6 & 5.33 & 64 \\
& & 18 & 15.42 & 185 \\
& & 36 & 32.63 & 391 \\
\cline{2-5}
& \multirow{3}{*}{$4$} & 6 & 9.72 & 116 \\
& & 18 & 31.95 & 383 \\
& & 36 & 56.13 & 674 \\
\hline
\hline
\end{tabular}
\centering\caption{\label{comp_t}Computational times: total elapsed times are reported in minutes, while the cost per iteration is measured in milliseconds per iteration (mpi).}
\end{table}}

\section{Noordin Top terrorist network analysis}
\label{sec:terrorist}
We analyze the terrorist dataset with information pertaining to $n=79$ terrorists (\emph{actors}) and their attendance behavior to $d=45$ events of various nature, such as trainings, operations, bombings, financial and logistics meetings, together with their affiliations to a number of organizations, associated with the leader of the Indonesian terrorist network Noordin Top \citep{everton2012disrupting}. Rather than leaving out the five lone wolf terrorists, we include them into the analysis.

We run our \texttt{manet} algorithm for 30,000 iterations with a generous burn-in window of 15,000, to ensure convergence.  {\color{black} Raftery and Lewis' diagnostic check from the \texttt{R} package \texttt{coda} \citep{codapack} supports this choice, by returning a suggested number of MCMC iterations ranging from 3500 to 8000. Convergence is further investigated and supported by the MCMC traceplots and via the Heidelberg and Welch's stationary test (with p-values above 0.50 for all the chains).}
Posterior quantities are computed on the samples after burn-in. These were not affected by label-switching and thus did not need any post-processing. The lowest computed DIC value for three possible values of $K=\{2,3,4\}$ corresponds to $\text{DIC}(2)=1822.93$, and we therefore select $K=2$ \emph{parent} clusters, corresponding to $K^{\star}=4$ \emph{heir} clusters.

The results are reported in Table \ref{post_index}. The first \emph{heir} cluster, identifying units belonging to no \emph{parent} cluster, contains 5 units who are the `lone wolves', i.e. the terrorists attending no event and who were discarded  from the analysis of \cite{aitkin2016statistical}. Only two units are allocated into the second \emph{heir} cluster: these two individuals are Noordin Top and Azhari Husin, the leader  and his main collaborator of the terrorists network, respectively. They form a separate cluster because of their peculiar behavior of participating to most of the 45 events, having the highest raw number of attendances, respectively 23 and 17, and being involved in many of the logistic, financial, and decision-making meetings. The third \emph{heir} cluster is formed by 6 individuals sharing the same pattern of attendances and, in particular, being terrorists affiliated to a specific sub-group called `KOMPAK'. Finally, in the fourth \emph{heir} cluster we find the rest of the terrorists such as trainees, henchmen, and religious leaders, who attend the 45 events with a pattern that is an overlap between the two \emph{parent} clusters. {\color{black} These results are found using a uniform prior allocation to clusters. The same allocation is robustly found also with a Dirichlet prior specification that discourages units to belong to too many clusters, i.e. by setting $a^{\star}_h = K^{\star}$ if $\sum_{h} u_h = 1$, and $a^{\star}_h=1$ otherwise.}

\begin{figure}[p!]
%\includegraphics[scale=0.65,center]{bipartite.pdf}
%\centering\caption{\label{bipart} Bipartite (two-mode) representation of Noordin Top terrorists network dataset. Each red square node is an event, with corresponding label, while each circle node is a terrorist (labelled with a progressive number). Color scheme for circle nodes reflects terrorists allocation into clusters obtained by our model \texttt{manet}: 2 \textbf{green} nodes for cluster $\boldsymbol{z}=(0,1)$; 6 \textbf{blue} nodes for cluster $\boldsymbol{z}=(1,0)$; 66 \textbf{light blue} nodes for multiple allocation cluster $\boldsymbol{z}=(1,1)$; \textbf{grey} nodes $\{75,76,77,78,79\}$ are the `lone wolves', attending no event.}
\includegraphics[scale=0.7,center]{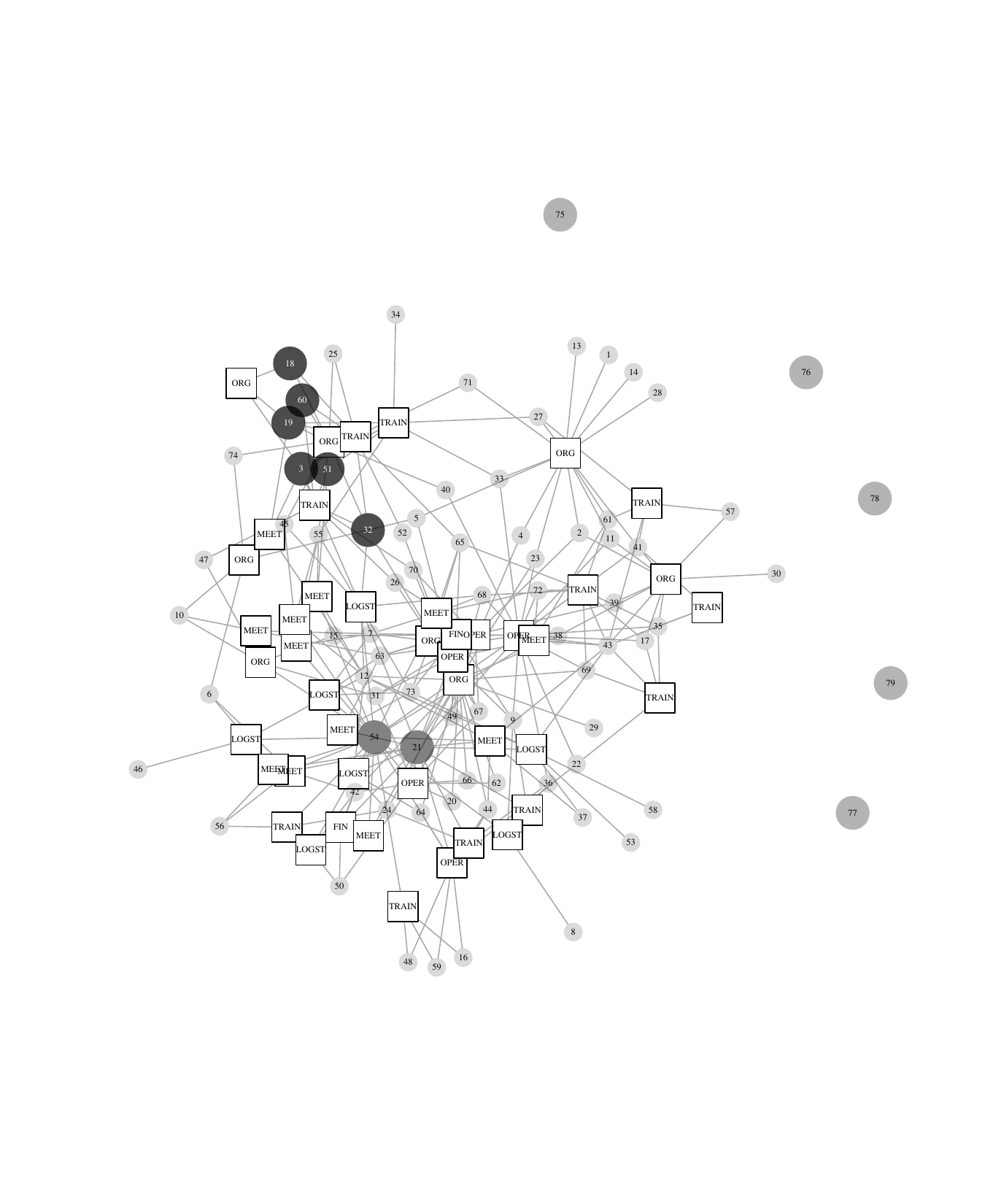} \vspace{-30mm}
\centering\caption{\label{bipart} Bipartite (two-mode) representation of Noordin Top terrorists network dataset. Each square node is an event, with corresponding label, while each circle node is a terrorist (labelled with a progressive number). Sizes and grey-shading scheme for circle nodes reflects terrorists allocation into clusters obtained by our model \texttt{manet}: 2 \textbf{medium shaded} nodes for cluster $\boldsymbol{z}=(0,1)$; 6 \textbf{heavy shaded} nodes for cluster $\boldsymbol{z}=(1,0)$; 66 \textbf{small sized, light shaded} nodes for multiple allocation cluster $\boldsymbol{z}=(1,1)$; \textbf{medium sized, medium shaded} unconnected nodes $\{75,76,77,78,79\}$ are the `lone wolves', attending no event.}

\end{figure}

\begin{figure}[ht!]
\centering
\includegraphics[scale=0.9,center]{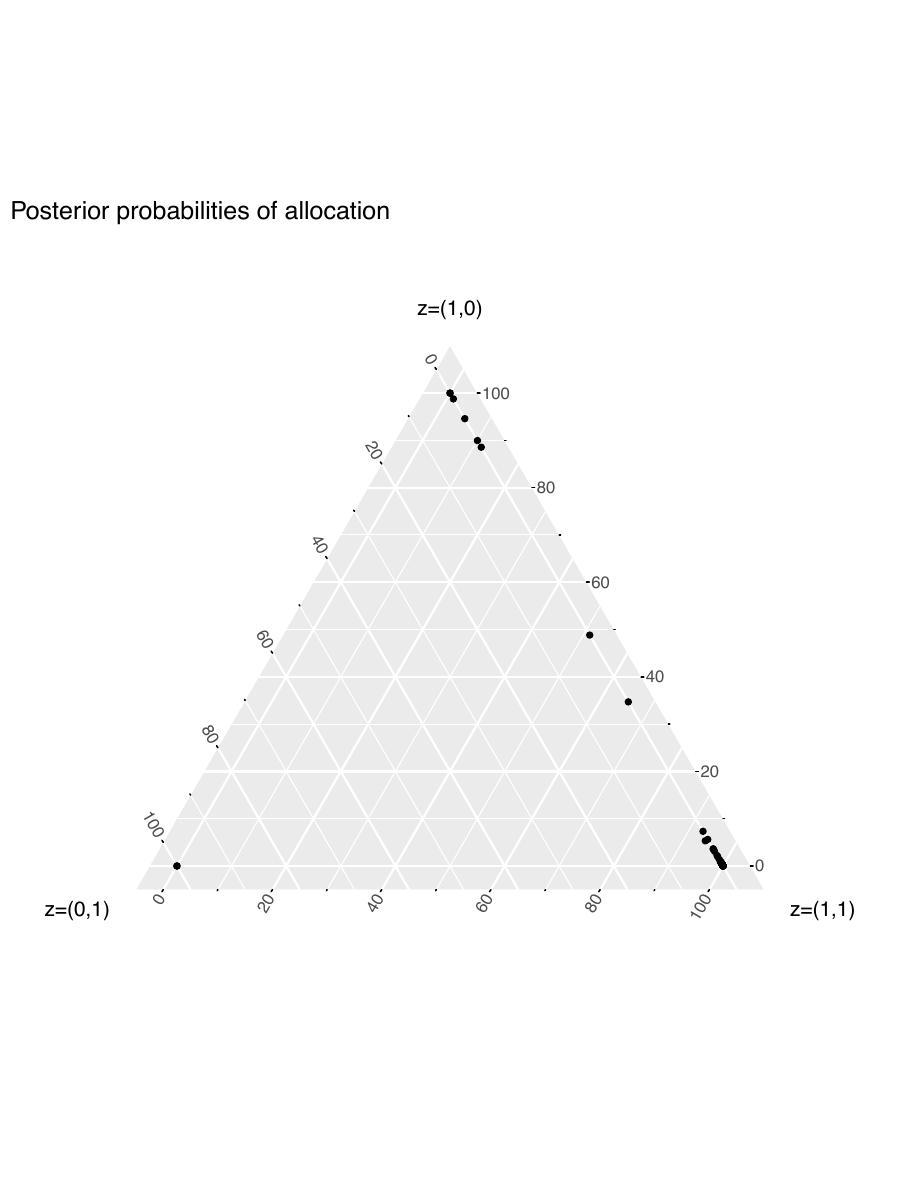}
\centering\caption{\label{tern_plot} Ternary plot for the (average) posterior probabilities of allocation of each terrorist to each clusters from our \texttt{manet} model, conditioning on not being in cluster $z=(0,0)$. The `lone wolves' cluster is omitted for ease of visualization.}
\end{figure}

Figure \ref{bipart} visualizes the two-mode (\emph{actor\textendash event}) Noordin Top network: red square vertexes are the events, with corresponding labeling; round vertexes are the terrorists, with a color scheme representation based on the clustering obtained with \texttt{manet}, and labelled with progressive numbers. Figure \ref{tern_plot} provides a graphical representation of the posterior probabilities averaged across the MCMC iterations (after burn-in). Each dot represents one of the 79 terrorists (the `lone wolves' are removed for visualization purposes): lower -- from left to right -- axis of the ternary plot depicts the posterior probability to be allocated into a multiple allocation cluster $\boldsymbol{z}_i=(1,1)$; similarly, the other two axes (left and right) measure the posterior probability to be allocated into cluster $\boldsymbol{z}_i=(0,1)$ -- top to bottom -- or cluster $\boldsymbol{z}_i=(1,0)$ -- bottom to top. We can see almost all units bear no uncertainty about their membership to the clusters, except for two terrorists, row 25 and 55 of the matrix. In order to report the uncertainty of the classification for all the groups, we provide the (PCM) in Table \ref{conf_mat}. As we see from the table, the results are close to a situation with no confusion in the classification except for cluster $\boldsymbol{z}_i=(0,0)$. This is partially expected because the data matrix is very rarefied and units in the multiple allocation cluster $\boldsymbol{z}_i=(1,1)$ attend very few events. This means that the attendance profile, and the cluster-specific vector of event probabilities $\boldsymbol{\pi}_h$, for cluster $h=1$ and $h=4$ are indeed very similar, pushing the algorithm to distinguish less the two groups. However, as we saw in Table \ref{post_index}, the `lone wolves' are classified into cluster $\boldsymbol{z}_i=(0,0)$, without any additional unit attending a low number of events.
\begin{table}[ht!]
\centering
\begin{tabular}{|c|cccc|}
\hline
\multicolumn{5}{|c|}{Rescaled PCM with $K=2$ ($K^{\star}=4$)} \\
\hline
Cluster & $z=(0,0)$ & $z=(0,1)$ & $z=(1,0)$ & $z=(1,1)$ \\
\hline
$z=(0,0)$ & 0.66 & 0.00 & 0.00 & 0.34 \\
$z=(0,1)$ & 0.00 & 1.00 & 0.00 & 0.00 \\
$z=(1,0)$ & 0.00 & 0.00 & 0.94 & 0.06 \\
$z=(1,1)$ & 0.01 & 0.00 & 0.01 & 0.98 \\
\hline
\hline
\end{tabular}
\centering\caption{\label{conf_mat}Rescaled posterior confusion matrix of the classification for 79 terrorists; the benchmark for comparison (best case scenario) is the identity matrix of order 4.}
\end{table}

For comparison, we explore results from our direct competitor \texttt{mixtbern} and consider both the case of $K=4$ and $K=8$ non-overlapping clusters. In both models, only three clusters are non-empty and the partitioning of the units into these mutually exclusive groups allocates Noordin Top and his main collaborator (Azhari Husin) into two separate singletons, whereas all the remaining terrorists are allocated into one of the other clusters.  Table \ref{tab_mixtbern4} reports the number of allocated units in each cluster, and the corresponding PCM, for the case $K=4$ (similar results for $K=8$ are reported {\color{black}in the Supplementary file}). The results suggest that allowing for and modelling the potential overlaps of the terrorists groups  in attending events, as is done in \texttt{manet}, helps in better identifying the subgroups in the network. In addition, we can find similarities and differences with the analysis in \citet{aitkin2016statistical}. Firstly, in both analyses, aside from the `lone wolves', data seem to point towards a 3-groups structure. Secondly, while the `lone wolves' are removed in the analysis of  \citet{aitkin2016statistical}, we are able to naturally account for terrorists belonging to the network but showing no attendances to the events considered. Finally, Azhari Husin and Noordin Top are allocated together into a two-units group in both analyses, but  terrorists' memberships to the other two remaining clusters are more confused in \citet{aitkin2016statistical} than with our model in terms of posterior allocations (see Figure 10 of their manuscript).

{\color{black}As a final analysis, given that the events have a natural grouping structure, we compare the full model with a collapsed version of \texttt{manet}, where columns - events - are gathered according to their nature (financial meetings, organizations, etc). In this case, the number of parameters is smaller than the original formulation, as we only have $\hat{d}=6$ groups of events instead of $d=45$. The lowest value for the DIC is obtained again with $K=2$, and it is equal to $\text{DIC}(2)=1884.34$. Comparing this with the earlier result (DIC(2) = 1822,93) suggests that the information about the grouping of the events, based on their category, is only partly explaining the clustering structure of the terrorists.}

\begin{table}[ht!]
\centering
\begin{tabular}{|c|c|c|l|}
\hline
\multicolumn{2}{|c|}{Clusters} & N. of individuals & \multirow{2}{*}{Qualitative Description}\\
\cline{1-2}
\emph{parent} cluster & \emph{heir} cluster & & \\
\hline
$z=(0,0)$ & $h=1$ & 5 & `lone wolves' \\
$z=(0,1)$ & $h=2$ & 2 & Noordin Top and Azhari Husin \\
$z=(1,0)$ & $h=3$ & 6 & KOMPAK sub-cell group \\
$z=(1,1)$ & $h=4$ & 66 & trainees and henchmen \\
\hline
& & 79 & \\
\hline
\end{tabular}
\centering\caption{\label{post_index}Posterior allocation of the 79 terrorists into $K^{\star}=4$ \emph{heir} clusters from our \texttt{manet} model, according to the MAP rule. First column shows the corresponding latent representation in the original parametrization.}
\end{table}

\begin{table}[ht!]
\centering
\begin{tabular}{|c|c|l|}
\hline
Clusters & N. of individuals & Qualitative Description\\
\hline
$k=1$ & 1 & Noordin Top\\
$k=2$ & 1 &  Azhari Husin \\
$k=3$ & 77 & all other terrorists \\
$k=4$ & 0 & - \\
\hline
& 79 & \\
\hline
\end{tabular}
\begin{tabular}{|c|cccc|}
\hline
\multicolumn{5}{|c|}{Rescaled PCM with $K=4$} \\
\hline
Cluster & $k=1$ & $k=2$ & $k=3$ & $k=4$ \\
\hline
$k=1$ & 1.00 & 0.00 & 0.00 & 0.00 \\
$k=2$ & 0.00 & 1.00 & 0.00 & 0.00 \\
$k=3$ & 0.00 & 0.00 & 1.00 & 0.00 \\
$k=4$ & 0.00 & 0.00 & 0.00 & 1.00 \\
\hline
\end{tabular}
\centering\caption{\label{tab_mixtbern4} Modelling the Noordin Top network using a non-overlapping mixture model (\texttt{mixtbern}) with $K=4$ clusters. (\emph{top}) Posterior allocation of the 79 terrorists into the 4 clusters according to MAP rule, (\emph{bottom}) Rescaled posterior confusion matrix of the cluster allocation for the 79 terrorists.}
\end{table}

\section{Conclusions}
\label{sec:Conclusions}
In this paper, we have presented a novel finite mixture model and have shown its applicability to the clustering of actor\textendash event data. We have formulated the model in a way that the actor\textendash event data can be modeled directly without transforming it to the more traditional actor\textendash actor network data, with the inherent loss of information. The general formulation of the model, with potentially overlapping clusters, allows for actors to belong to multiple communities on the basis of their pattern of attendances to events. The model itself allows to define the meaning of overlap, leading to a reduction in the number of parameters as well as a clearer interpretation of the results.

Using our model on the Noordin Top actor\textendash event network, we discovered three distinct subgroups out of the 79 terrorists on the basis of their mode of attendance to 45 meetings: the first group consisted of 5 suicide bombers who did not attend any meeting, the second group consisted of 6 members of the KOMPAK terrorist organization and the third group consisted of the 2 leaders, namely Top and Husin. This view of the terrorist network gives a more layered understanding of the mode of operation and allegiances within the organization.

We proposed a Bayesian inference procedure for deriving the posterior distribution of the parameters in the model. By selecting appropriate conjugate prior distributions, the MCMC sampler is efficient and convergence is typically fast. The proposed model is currently implemented in the \texttt{R} package \texttt{manet}, available on \texttt{CRAN}. The package contains the Noordin Top terrorist network used for this paper, as well as the Southern US Mississippi women dataset and the larger synthetic dataset discussed in the {\color{black}Supplementary Material.}

The Bayesian formulation of the model lends itself naturally to an extension of the model to include also individual level covariates, either at the level of group membership or event attendance probabilities. {\color{black} This would on the one hand adjust for node degree/hetereogeneity and on the other hand enhance the interpretability of the resulting clusters. In applications where the second mode does not have a known grouping structure, as it was the case for the Noordin Top network and the grouping of events, future work will develop extensions to bi-clustering with overlap. Finally, possible extensions could consider introducing dependency among events, thus relaxing the local independence assumption currently used, and addressing the case of weighted and dynamic networks.}

\section*{Acknowledgements}
The authors would like to acknowledge the contribution of the COST Action CA15109 (COSTNET), which funded a visit of the first author to Brunel University London.

\bibliographystyle{imsart-nameyear}
\bibliography{biblio_terr}

%%%%%%%%%%%%%% TABLES AND FIGURES

\end{document}